# Direct Measurement of Folding Angle and Strain Vector in Atomically thin $WS_2$ using Second Harmonic Generation

Ahmed Raza Khan[1], Boqing Liu[1], Wendi Ma[1], Linglong Zhang[1], Ankur Sharma[1], Yi Zhu[1], Tieyu Lü[2] and Yuerui Lu[1*]

[1]Research School of Electrical, Energy and Materials Engineering, College of Engineering and Computer Science, Australian National University, Canberra ACT, 2601, Australia

[2]Department of Physics, and Institute of Theoretical Physics and Astrophysics, Xiamen University, Xiamen, 361005, China

**\*** To whom correspondence should be addressed: Yuerui Lu (yuerui.lu@anu.edu.au)

## ABSTRACT

Structural engineering techniques such as local strain engineering and folding provide functional control over critical optoelectronic properties of 2D materials. Accurate monitoring of local strain vector (both strain amplitude and direction) and folding angle in 2D materials is important to optimize the device performance. Conventionally, the accurate measurement of both strain amplitude and direction requires the combined usage of multiple tools, such as atomic force microscopy (AFM), electron microscopy, Raman spectroscopy, etc. Here, we demonstrated the usage of a single tool, polarization-dependent second harmonic generation (SHG) imaging, to determine the folding angle and strain vector accurately in atomically thin tungsten disulfide ($WS_2$). We find that trilayer $WS_2$ folds with folding angle of $60^0$ show 9 times SHG enhancement due to vector superposition of SH wave vectors coming from the individual folding layers. Strain dependent SHG quenching and enhancement is found parallel and perpendicular respectively to the direction of the compressive strain vector. However, despite a variation in strain angle, the total SHG remains constant which allows us to determine the local strain vector accurately using photoelastic approach. We also demonstrate that band-nesting induced transition (C peak) can highly enhance SHG, which can be significantly modulated by strain. Our results would pave the way to enable novel applications of the TMDs in nonlinear optical device.

Two-dimensional (2D) layered semiconductor materials such as transition metal dichalcogenides (TMDs) have received tremendous attention due to their interesting optoelectronic properties and potential applications in electronic devices.[1–3] Tuning the optoelectronic properties of these materials is important for the optimum device performance.[4] Therefore, researchers have used various ways to tune the properties of 2D materials such as structural engineering[5,6], defect engineering[7], doping[8], etc.

Structural engineering of 2D materials provide an exciting platform to tailor the material's properties through modification in lattice structure. For example, 2D Graphene sheet, a zero bandgap structure, is rolled to form carbon nanotubes with tunable bandgap depending on rolling angle. Armchair nanotubes are metallic structures whereas zigzag nanotubes show semiconducting properties with open bandgap.[9–11] Modulation in electronic structure and PL properties is reported through twisting angle modification in TMDs heterostructures.[12–14]

Strain engineering[15] and folding[16] are two important types of structural engineering techniques to tune optoelectronic properties. For instance, strain engineering is shown to reduce the carrier effective mass and modify the valley structure of atomically thin $MoS_2$, thus leading to an increase in its career mobility.[17–24] In addition, significant photoluminescence (PL) enhancement is reported in strained atomically thin $WSe_2$ due to bandgap modulation.[25] Similarly, folded structures of $MoS_2$ are reported to tune PL intensity due to modulation in interlayer coupling.[26] Folding angle modulation in $MoS_2$ is shown to tailor electron and phonon properties.[27] Because both strain engineering and folding provide an effective way to tune optoelectronic properties and improve the performance of optoelectronic devices, therefore, there is a need of full assessment of local strain vector and folding parameters to utilize their full potential.

Conventionally, determination of both strain amplitude and direction requires combination of multiple tools. For instance, researchers use atomic force microscopy (AFM) to measure the strain amplitude on strain induced wrinkles[6] whereas electron/neutron microscopy is used to determine the relation of strain direction to lattice structure.[28] Recently, optical second-harmonic generation (SHG) has been shown to probe the crystallographic orientation, lattice symmetry and stacking order of non-inversion symmetric 2D materials such as odd layers of TMDs, hBN, Group IV monochalcogenides, etc.[29–32] Because SHG intensity is very sensitive to the structural configurations of 2D materials; it is, in principle, feasible to employ SHG to monitor folding and straining in 2D materials.

Here, we have used polarization-dependent SHG as a single tool to probe folding angle and strain vector precisely in atomically thin tungsten disulfide ($WS_2$). Trilayer folds with $60°$ folding angle are found to show 9 times SHG enhancement due to the vector superposition of SH wave vectors coming from the individual layers of the folds. We find strain dependent SHG quenching and enhancement, parallel and perpendicular respectively to the direction of the compressive strain vector. However, strain angle dependent total SHG (without polarizer) remains constant which allows us to find the local strain vector accurately using photoelastic effect. We find SHG to be very sensitive to C-exciton can be tuned through strain modification. Our results show SHG as a powerful tool to probe both folding angle and strain vector in atomically thin TMDs.

## Results

**Differentiation of wrinkles and folds by SHG**

In this work, we have used mechanical buckling of the flexible substrate to obtain folds[16] (1-3L) and strained wrinkles[6] (5-6L) in atomically thin $WS_2$. The details of the fabrication method are shown in **Figure 1a** and given in methods section and **S1-S2** in supplementary information. Optical microscopic images of folds (1-3L) are shown in **Figure 1b**. Phase Shifting Interferometry (PSI) is employed to identify the layer number.[33–37] We have used 900nm laser excitation confocal light microscope for second harmonic generation (SHG) mapping (450nm) of flat and folded regions of 1-3L $WS_2$ as shown in **Figure 1c** (see methods section for more details).

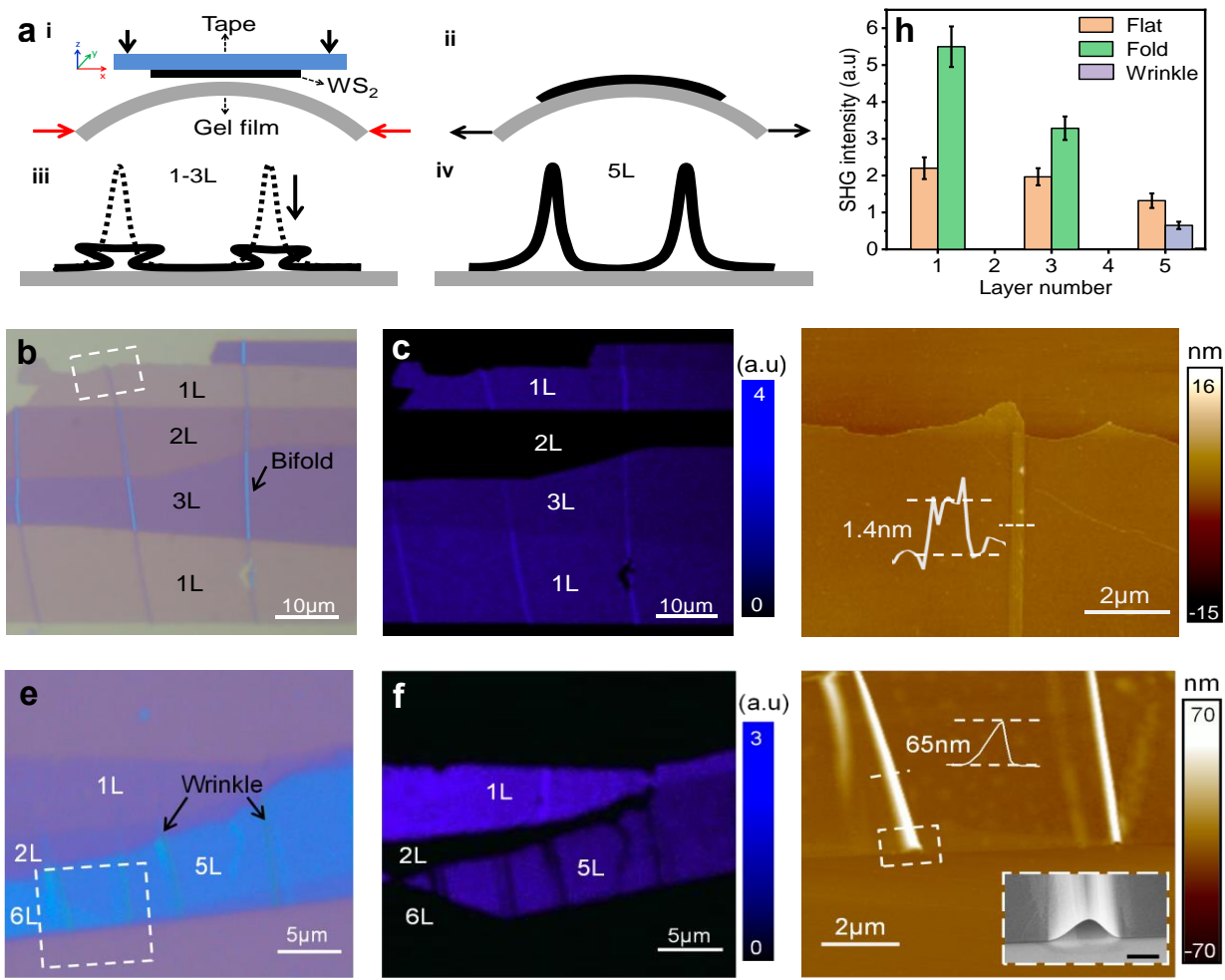

**Figure 1 | Differentiation of wrinkle and fold nano-structures by SHG (a)** Schematic diagram of the fabrication process of buckled $WS_2$ sample. **(b)** Optical microscopic image of 1-3L $WS_2$ sample fabricated by the process described in (a), showing the formation of folds due to collapse of wrinkles. **(c)** AFM (Atomic force microscopy) topography image of the region marked by the white dashed rectangle in (b) **(d)** Optical microscopic image of 5-6L buckled $WS_2$ sample showing strained wrinkles on 5L and 6L. **(e)** SHG intensity mapping of

the region shown in **Figure 1(b)**. The mapping shows SHG enhancement on 1L and 3L folded regions. **(f)** SHG intensity mapping of the region shown in **Figure 1(e)**. The SHG mapping shows reduction in SHG on 5L wrinkles. **(g)** AFM topography image of the region marked by the white dashed rectangle in (e). Inset shows SEM (Scanning Electron Microscopy) of the wrinkle's profile **(h)** A stat-plot showing the SHG response for flat, folded (1L & 3L) and strained wrinkled (5L) regions for ultrathin $WS_2$. Histogram shows the SHG intensity response, with uncertainties indicated by the error bars. The light brown, green and light blue rectangles indicate the SHG intensity measurements for flat, folded and wrinkled regions respectively. All the measurements are taken at 900nm laser excitation.

Odd layers i.e 1L, 3L and 5L show SHG signal due to non-centrosymmetric structure whereas even layer numbers do not show SHG signal due to centrosymmetric structure which is consistent with the previous studies.[32] Interestingly, a significant higher SHG response (~2-3 times) from folded regions is observed as compared to flat regions as shown in **Figure 1c**. Power dependent SHG on flat, folded and wrinkles regions is performed to confirm if the photons collected are SH photons. The corresponding SHG signal intensity is drawn with excitation power on a log scale. A fitted value ~ 2 on logscale for power vs SHG intensity confirms the collected photons as SH photons[38–40] (**Figure S2**). Atomic Force Microscopic (AFM) investigation shows that the height differences measured on the 1L, 2L and 3L folds of $WS_2$ are found to be 1.4 ± 0.5, 2.8 ± 0.5 and 4.2 ± 1 nm respectively (**Figure 1d and Figure S3**). These values match the height of 2L, 4L and 6L $WS_2$ very well as the thickness of single layer is evaluated around 0.7 nm[41], which confirms the bifold formation (such as trilayer fold or 1L+1L+1L on 1L $WS_2$) in 1-3L $WS_2$. SHG investigation of 5L wrinkles shows a drop in SHG as compared to flat 5L (**Figure 1f**) which will be explained later. AFM investigation of 4-6L wrinkles reveals a rapid increase in the height (~50-70nm) as shown in **Figure 1g and Figure S3**. The wrinkle like curvature in Scanning electron microscopy (SEM) examination confirms that wrinkles maintain their curvature in ≥4L in $WS_2$. **(Figure 1g)**

**Folds SHG**

In the previous section, we showed SHG enhancement on folds. The SH response from the fold can be modeled by the vector superposition of all the layers of the fold which is explained here. Let's consider the case of trilayer fold (1L+1L+1L) on 1L WS$_2$. Opening up of 1L fold shows that the top layer of the fold (designated as L$_1$ in **Figure 2a**) is parallel to the bottom layer (L$_3$), which implies arm chair direction of L$_1$ (shown as the black line bisecting the hexagonal WS$_2$ and black triangle in L$_1$) is parallel to the armchair direction of L$_3$ (green bisecting line), whereas armchair direction of the mid layer L$_2$ (blue bisecting line) of the fold makes an angle of 180° with the arm chair direction of L$_1$ and L$_3$. 1L WS$_2$ belongs to D$_{3h}$ symmetry, therefore, it shows a six-fold polar SH response as under[32];

$$I_{//}(2\omega) \propto cos^2 3\varphi \qquad (1)$$

where $I_{//}(2\omega)$ is the SHG intensity for parallel polarization (i.e polarizer is parallel to the direction of polarization component of incident laser) and $\varphi$ is the azimuthal angle between the polarized incident laser and the armchair direction.[32,42] SHG intensity becomes maximum when the incident laser polarization is parallel to the armchair direction.[32] For our folding case, $I_{L1} = I_{L3} \propto cos^2 3\varphi_1$ and $I_{L2} \propto cos^2 3(\varphi_1+180+\theta_f)$ where $I_{L1}(2\omega)$, $I_{L2}(2\omega)$, and $I_{L3}(2\omega)$ are the SHG wave vector responses from L$_1$, L$_2$ and L$_3$ of the fold. Hsu. et. al.[29] reported that SH wave vector from two stacked layers ($I_s$) under parallel polarization can be found by the vector superposition of SH wave vectors from two individual layers as under;

$$I_{s_{//}}(2\omega) \propto I_a + I_b + 2\sqrt{(I_a * I_b)}\, cos3(\theta) \qquad (2)$$

where θ is the stacking angle between the armchair directions of *a* and *b*. Thus, SHG response from the fold can be solved by the vector superposition of the SHG response coming from the individual layers of the fold i.e $I_{L1}(2\omega)$, $I_{L2}(2\omega)$ and $I_{L3}(2\omega)$ as shown in **Figure 2b**.

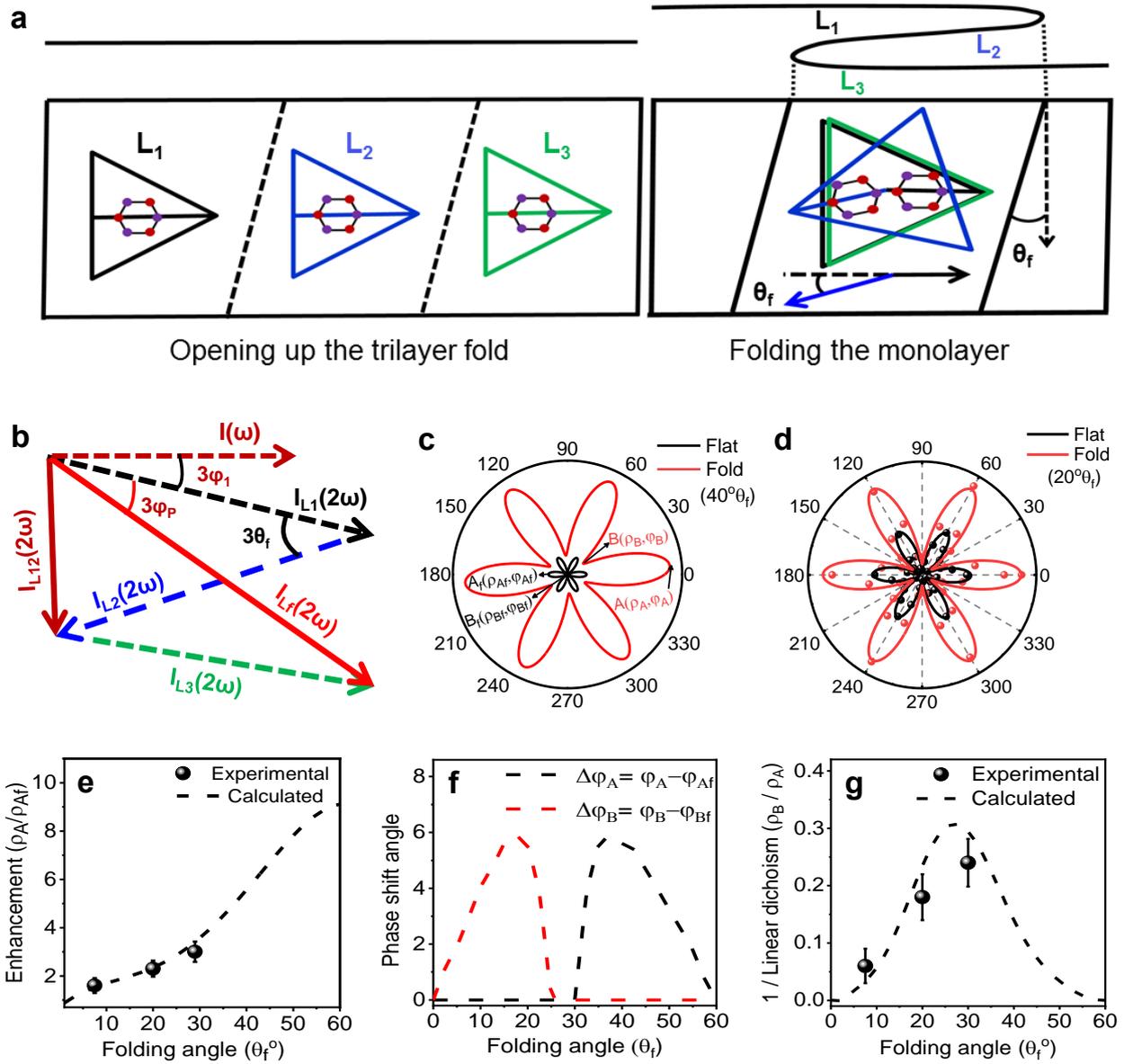

**Figure 2 | Engineering SHG through folding of atomically thin TMDs. (a)** A schematic illustration for the stacking of layers in a trilayer fold (*1L+1L+1L*). The top (layer 1) and bottom layer (layer 3) are parallel to each other, whereas armchair direction of mid layer (layer 2) makes an angle of (180+$\theta_f$) with the armchair direction of top and bottom layer. [The lines (black, blue and green) bisecting the triangles (black, blue and green) show the armchair direction] **(b)** Vector superposition of the SH fields from the layer of the fold, where $I_{L1}(2\omega)$ (black line), $I_{L2}(2\omega)$ (blue line) and $\mathbf{I}_{L3}(2\omega)$ (green line) are the SH wave vectors from $L_1$, $L_2$ and $L_3$ respectively, $I(\omega)$ (brown line) is the laser wave vector and $I_{Lf}(2\omega)$ is the resultant SH wave vector from the fold. $3\varphi_1$ is the phase shifting angle between input linearly polarized laser and $I_{L1}(2\omega)$ whereas $3\varphi_P$ is the phase shifting angle between $I_{L1}(2\omega)$ and

$I_{Lf}(2\omega)$. **(c)** Calculated SHG $I_{ll}(2\omega)$ polar response for 1L flat and fold (40° $\theta_f$). A($\rho_A$, $\varphi_A$) is the maximum SHG ($\rho_A$) amplitude point for the fold with $\varphi_A$ (degrees) angle from 0° whereas B($\rho_B$, $\varphi_B$) is the minimum SHG ($\rho_B$) amplitude point for the fold with $\varphi_B$ (degrees) angle from 0°. A$_f$ ($\rho_{Af}$, $\varphi_{Af}$) and B$_f$ ($\rho_{Bf}$, $\varphi_{Bf}$) represent the maximum and minimum point of the flat region. **(d)** Experimental investigation of polarization resolved SHG $I_{\parallel}(2\omega)$ intensity pattern for 1L flat and folded WS$_2$. Continuous lines are the fitted plots, whereas symbols are experimental data points. **(e)** The folding angle dependence of SHG phase shift angle (degrees) **(f)** The folding angle dependence of SHG enhancement for fold, where enhancement= $\rho_A$ / $\rho_{Af}$. **(g)** The folding angle dependence of (Linear dichrisom)$^{-1}$ where (Linear dichrisom)$^{-1}$ = $\rho_B$ / $\rho_A$. Dashed line is the calculated response whereas spherical symbols are the experimental data points. Error bars represent the range of error in the measured values.

In case of our trilayer fold, this can be done by the vector addition of two entities first ($I_{L1}(2\omega)$ and $I_{L2}(2\omega)$) to find their resultant $I_{L12}(2\omega)$ where $\theta = 180+\theta_f$ and then adding this resultant vector $I_{L12}(2\omega)$ to the third entity vector ($I_{L3}(2\omega)$) to get the overall resultant vector $I_{Lf}(2\omega)$ where $I_{Lf}(2\omega)$ is SH wave vector from the fold. $\varphi_I$ is the azimuthal angle between incident laser polarization component and armchair direction of $I_{L1}(2\omega)$ whereas $3\varphi_P$ is the phase shifting angle between $I_{L1}(2\omega)$ and $I_{Lf}(2\omega)$ as demonstrated in **Figure 2b**. Using the above scheme, the angular SHG response of folded region [$I_{Lf}(2\omega)$] with $\theta_f = 40°$ is calculated as shown in **Figure 2c** where $A(\rho_A, \varphi_A)$ is the maximum amplitude point of SHG ($\rho_A$) for the fold with $\varphi_A$ (degrees) angle from horizontal (0°) whereas $B(\rho_B, \varphi_B)$ is the minimum SHG ($\rho_B$) amplitude point for the fold with $\varphi_B$ (degrees) angle from horizontal. $A_f(\rho_{Af}, \varphi_{Af})$ and $B_f(\rho_{Bf}, \varphi_{Bf})$ represent the maximum and minimum points of the flat region. Here, $\varphi_{Af}$ and $\varphi_{Bf}$ represent AC (arm chair) direction at 0° and ZZ (zigzag) direction at 30° because we are using parallel polarization for SHG.

In order to experimentally investigate the polarization dependent SHG response of folded region, we put a polarizer in between sample and spectrometer in such a position that the polarization component of the SH radiation is parallel to the polarization state of the incident laser (900nm)i.e parallel polarization of SHG (see methods section for more details). We get an enhanced (~2.6) SHG polar response from the folded region ($\theta_f = 20°$) along the armchair direction as demonstrated in **Figure 2d** (See supplementary section **S4** for folding angle determination). As folding angle is expected to tune SHG intensity coming from the fold, we calculate SHG enhancement factor = $\rho_A /\rho_{Af}$ as indicated by the dashed line as shown in **Figure 2e.** The calculated angular SHG response shows 1 to 9 times SHG enhancement as $\theta_f$ goes from 0° to 60°. The experimental results are found in good agreement with the calculated values which shows the validity of our model predictions. Phase shifting angle of folded region is the angular variation in waveform of folded region w.r.t flat region. This measurement can be important in order to optimize the device performance. We, therefore, calculate the phase shifting angles as follow; (i) $\Delta\varphi_A = \varphi_A - \varphi_{Af}$ and (ii) $\Delta\varphi_B = \varphi_B - \varphi_{Bf}$. A maximum phase shift of 6° is found at 20° and 40° for $\Delta\varphi_B$ and $\Delta\varphi_B$ respectively as shown in **Figure 2f.** However, phase shifting angles are too small to be detected accurately within the resolution limit of our experimental setup. An anisotropy response of SHG intensity is expected to be influenced by folding angle, therefore, we are interested to calculate (linear dichroism (LD))$^{-1}$ = $\rho_B / \rho_A$ which shows a maximum value of 0.3 at 30° $\theta_f$ as shown in **Figure 2g.** Experimental investigation shows good agreement with the model prediction. The above results thus establish SHG as a powerful technique to monitor folds in atomically thin WS$_2$.

**Strain vector determination through SHG**

In the previous section, we showed that the wrinkles on 5L do not collapse and maintain their wrinkles' like curvature; therefore, SHG response of wrinkles is expected to be influenced by the local strain vector. In this context, we run polarization-dependent SHG on the flat and winkled regions (P$_1$ and P$_2$) of 5L WS$_2$ (**Figure 3a-3c**) using pump 830nm laser which is initially aligned with the armchair direction of the flat region. Similar to 1L, we get a uniform six fold SHG polar pattern from flat 5L WS$_2$ due to D$_{3h}$ symmetry (**Figure 3a**).

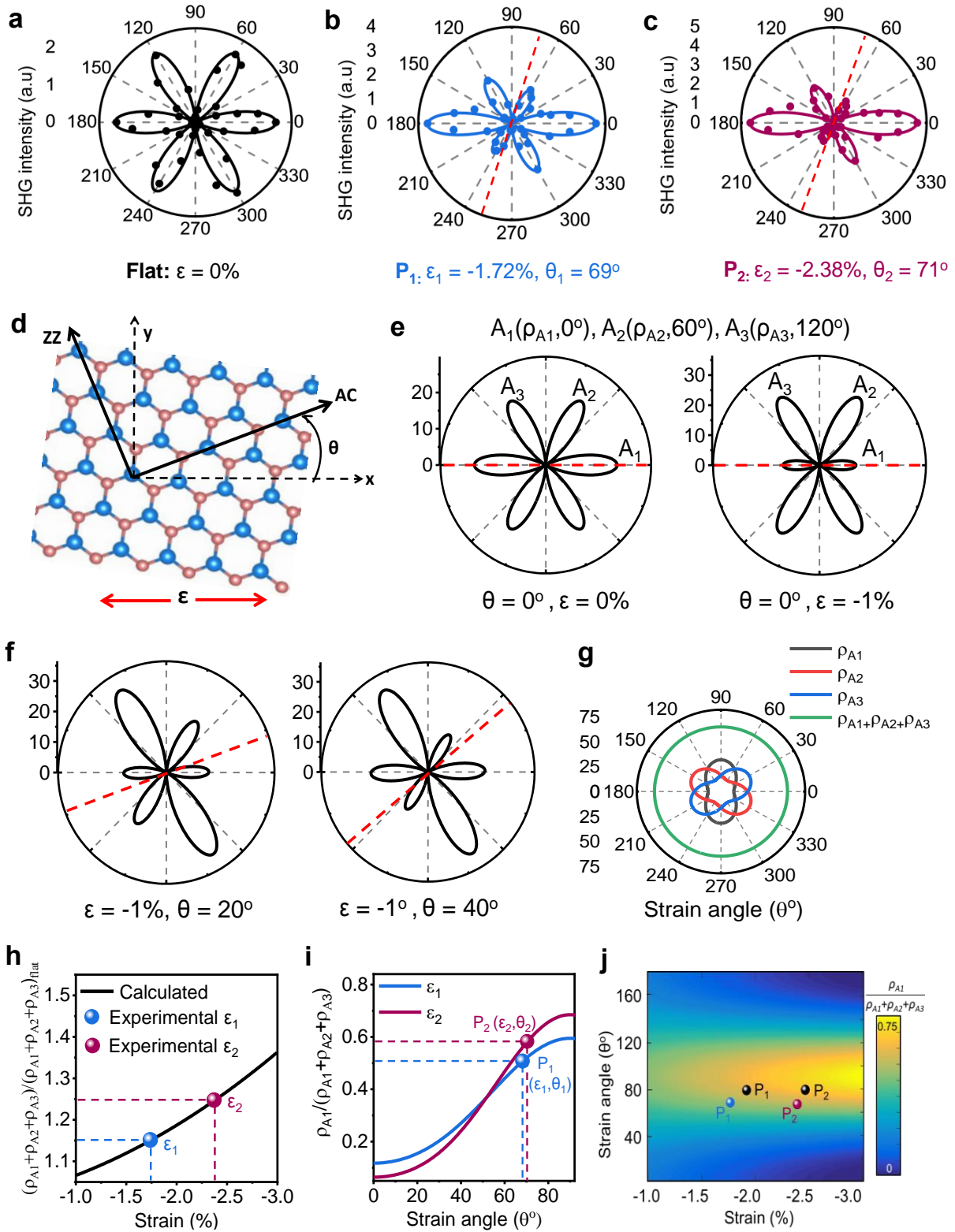

**Figure 3 | Determination of strain vector by angle resolve SHG** (a) Experimental investigation of polarization resolved SHG intensity $I_{ll}(2\omega)$ pattern for for flat (b) $P_1$ wrinkle (c) $P_2$ wrinkle of 5L $WS_2$. Continuous lines are the fitted plots, whereas symbols are experimental data points. Red dashed line indicates the direction of strain. Determination of strain amplitude and direction is given in the following figures. (d) A schematic illustration to show the uniaxial strain $\varepsilon$ applied along the horizontal direction with $\theta^o$ strain angle between strain direction and AC (arm chair) direction. Tensile strain and compressive strain are indicated by positive and negative signs respectively. ZZ: zigzag. (e) Calculated polar response of SHG intensity $I_{ll}(2\omega)$ for $WS_2$ at the strain levels of 0% and -1% ($0^o$ strain angle). $A_1(\rho_{A1},0^o)$, $A_2(\rho_{A2},60^o)$ and $A_3(\rho_{A3},120^o)$ are three points with SHG intensity of $\rho_{A1}, \rho_{A2}$ and $\rho_{A3}$ in $0^o$, $60^o$ and $120^o$ direction with respect to horizontal direction. SHG intensity reduces and increases in the direction along and perpendicular to the compressive strain vector. (f) Calculated SHG $I_{ll}(2\omega)$ polar response of $WS_2$ at strain angles of $20^o$ and $40^o$ ($\varepsilon = -1\%$). (g) Calculated strain angle dependent $\rho_{A1}, \rho_{A2}, \rho_{A3}$ and $\rho_{A1}+\rho_{A2}+\rho_{A3}$ ($\varepsilon = -1\%$) (h) Strain dependent calculated ratio of $\rho_{A1}+\rho_{A2}+\rho_{A3}$ and $(\rho_{A1}+\rho_{A2}+\rho_{A3})_{flat}$ i.e $(\rho_{A1}+\rho_{A2}+\rho_{A3}) / (\rho_{A1}+\rho_{A2}+\rho_{A3})_{flat}$, $\varepsilon_1 = -1.72\%$ (blue sphere) and $\varepsilon_2 = -2.38\%$ (plum sphere) represent two strain values extracted from $P_1$ and $P_2$ wrinkles' SHG polar plots shown in Figure 3a-c where $(\rho_{A1}+\rho_{A2}+\rho_{A3}) / (\rho_{A1}+\rho_{A2}+\rho_{A3})_{flat} = 1.15$ and 1.25 for $P_1$ and $P_2$. (i) Strain angle ($\theta^o$) dependent calculated $\rho_{A1} / (\rho_{A1}+\rho_{A2}+\rho_{A3})$ for $\varepsilon_1$ (blue line) and $\varepsilon_2$ (plum line). $\theta_1 = 69^o$ (blue sphere) and $\theta_2 = 71^o$ (plum sphere) represent strain angles extracted from $P_1$ and $P_2$ SHG polar plots where $\rho_{A1} / (\rho_{A1}+\rho_{A2}+\rho_{A3}) = 0.51$ ($P_1$) and 0.58 ($P_2$). (j) SHG contour map showing strain (%) and strain angle ($\theta^o$) of $P_1$ (blue sphere) and $P_2$ (plum sphere) wrinkles extracted from SHG polar plots solely. The measured strain and strain angle using combination of AFM and SHG are displayed as black spheres for the comparison.

On the other hand, we get a distortion in SHG polar pattern on strain induced wrinkles ($P_1$ and $P_2$) as shown in **Figure 3b-3c.** In order to understand SHG polar pattern evolution for strain induced wrinkles, we consider photoelastic effect, an established approach,[43] for the explanation of SHG polar pattern evolution under strain. Let's take the case of odd layer $WS_2$ ($D_{3h}$ symmetry) under uniaxial strain $\varepsilon$ along the horizontal direction as depicted in **Figure 3d**.

Strain angle ($\theta^o$) is defined as the angle between strain direction and AC (arm chair) direction. The parallel polarized SHG intensity "$I_{//}(2\omega)$" under uniaxial strain $\varepsilon$ for $D_{3h}$ symmetry class considering photoelastic effect, is;[43]

$$I_{//}(2\omega) \propto \frac{1}{4}(A\cos(3\varphi) + B\cos(2\theta + \varphi))^2 \quad (3)$$

where $A = (1 - v)(p_1 + p_2)(\varepsilon_{xx} + \varepsilon_{yy}) + 2\chi o$ and $B = (1+ v)(p_1 - p_2)(\varepsilon_{xx} - \varepsilon_{yy})$, $p_1$ and $p_2$ are the photoelastic cofficients, $\varepsilon_{xx}$ and $\varepsilon_{yy}$ are the values of the strain(%) along x and y direction where tensile strain and compressive values are taken as positive and negative respectively, $v$ is the poisson ratio, $\varphi$ is the polarization angle and $\chi_0$ is the nonlinear susceptibility parameter of the unstrained crystal lattice. Because the wrinkles' formation occurs due to the inward compressive forces, therefore, the polar SHG response is calculated for 0% and -1% strain amplitude (at $\theta=0^o$) using Poisson ratio of $v_{(WS2)} = 0.22$[44], $p_1= 0.75$ nm/V/%[45], $p_2 = -0.97$ nm/V/%[45] and $\chi_0 = 4.5$ nm/V[46] (**Figure 3e**). SHG quenching and enhancement is found in the direction parallel and perpendicular respectively to the strain direction. The quenching and enhancement in their respective directions increase with the strain amplitude. The pattern's distortion is associated with the direction of compressive strain vector as demonstrated for variable strain angles (20°, 40°) in **Figure 3f**. If $A_1(\rho_{A1},0^o)$, $A_2(\rho_{A2},60^o)$ and $A_3(\rho_{A3},120^o)$ represent three points with SHG intensity of $\rho_{A1}$, $\rho_{A2}$ and $\rho_{A3}$ along three AC directions $\varphi = 0^o$, $60^o$ and $120^o$ respectively (Figure 3e), we find an invariable total SHG intensity ($\rho_{A1}+\rho_{A2}+\rho_{A3}$) irrespective of the strain angle (**Figure 3g**). As the total SHG intensity ($\rho_{A1}+\rho_{A2}+\rho_{A3}$) remains constant for each value of strain angle, we can use this finding to calculate strain dependent $(\rho_{A1}+\rho_{A2}+\rho_{A3})/(\rho_{A1}+\rho_{A2}+\rho_{A3})_{flat}$ to extract $\varepsilon_1 = -1.72\%$ and $\varepsilon_2 = -2.38\%$ from SHG polar plots of $P_1$ and $P_2$ wrinkles (**Figure 3h**). Hereafter, we calculate **s**train angle dependent $\rho_{A1}/(\rho_{A1}+\rho_{A2}+\rho_{A3})$ for $\varepsilon_1$ and $\varepsilon_2$ values as shown in **Figure 3i** to extract $\theta_1 = 69^o$ and $\theta_2 = 71^o$ from SHG polar plots. The extracted values are displayed in a contour plot (**Figure**

**3j**). Extracted values of strain angle are very close to each other because the wrinkles (on the same sample) are parallel to each other. Using the extracted values of strain vector, we find that SHG polar plots (Figures 3b-3c) fit well to photoelastic behaviour and degree of distortion relates well with the extracted values of strain vector. In order to validate our model calculations, we use AFM and polarization dependent SHG to measure the strain $\varepsilon_{xx}$ (%) and the strain angle ($\theta^o$) respectively for the selected wrinkles (See section S6 and S7 for more details). The measured values (black spheres in **Figure 3j**) show good agreement with the extracted measurements from SHG polar plots using the scheme given in Figure 3a-3i. Conventionally, the strain direction and amplitude measurements require the combination of multiple tools, such as AFM and polarization dependent SHG are required for these measurements in our case. However, SHG as a single powerful tool has the potential to probe strain amplitude and direction in 2D materials.

**Folds and wrinkles Investigation through wavelength dependent SHG**

SHG response is expected to be influenced by variation in laser wavelength, therefore, we run laser excitation wavelength dependence (810≤λ≤950nm) on the flat, folded and wrinkled regions of ultrathin WS$_2$ (**Figure 4a-c**). Here, we find SHG peak centered at 870-880nm laser wavelength (or 435-440nm of SHG wavelength) in wavelength dependent SHG  For 1L flat region (**Figure 4b**). The peak position is attributed to the resonance phenomenon due to the presence of C-exciton.[38,47] (**Figure 4d**). Wavelength dependent SHG response of 1L folds shows SHG enhancement at all the wavelengths scanned (**Figure 4b**). Recently, SHG is found highly sensitive towards strain according to the recent research.[48] This sensitive behaviour is shown by significant (49%) SHG quenching per 1% strain in atomically thin MoSe$_2$.[48] Such highly sensitive behaviour is also reported for MoS$_2$.[43] However, no study is reported on the origin of such sensitivity of SHG towards strain in 2D TMDs. Wavelength dependent SHG of strain induced 5L wrinkles show a remarkable blue shift in exciton resonance SHG peak (**Figure 4c**). This shift in SHG peak position appears to be dependent on strain amplitude. The shift in SHG peak position is attributed to the strain dependent shift in C-exciton resonance which is consistent with the literature.[49]

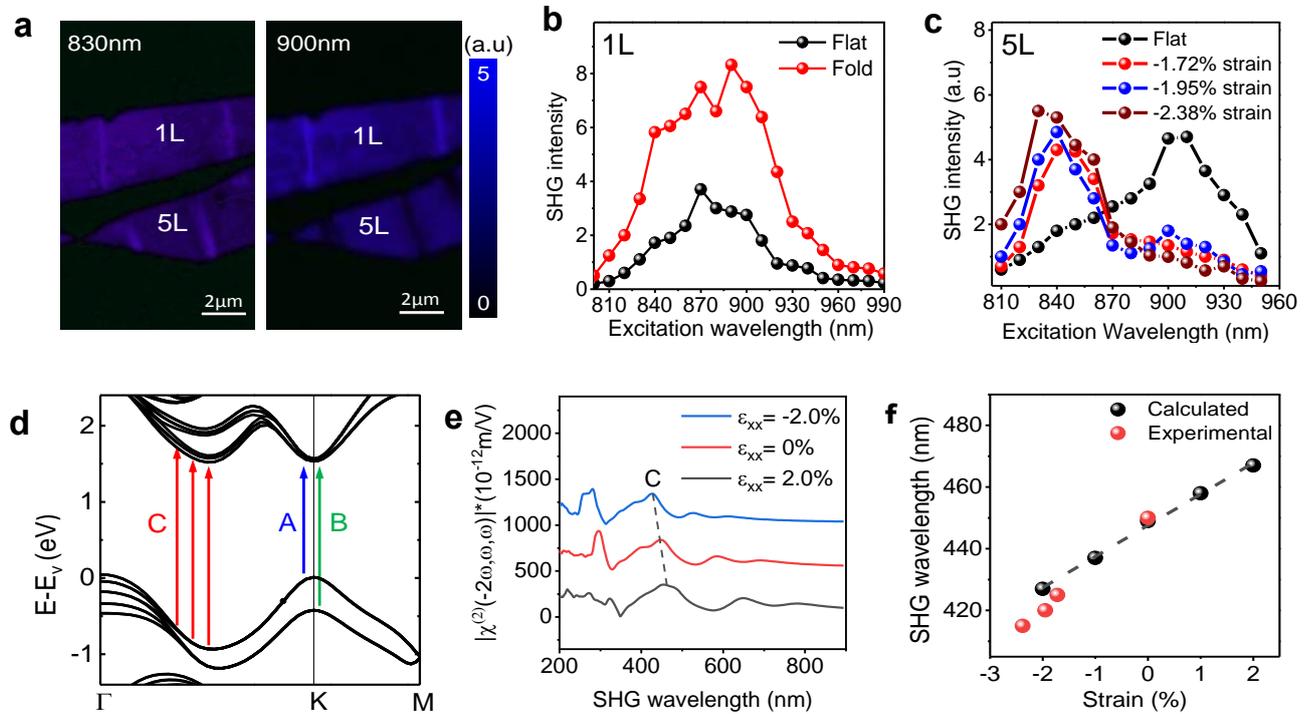

**Figure 4 | Differentiation of folds and wrinkles through wavelength dependent SHG** **(a)** SHG intensity mapping of 1L folds and 5L wrinkles for 830 nm and 900 nm. 1L folds show an enhancement on both the excitation wavelengths scanned whereas $WS_2$ wrinkles show an enhancement and reduction in SHG intensity at 830 nm and 900nm respectively. **(b)** Excitation wavelength dependent SHG intensity of folded and flat regions of 1L $WS_2$. Folds show an enhancement at all the wavelengths scanned. **(c)** Excitation wavelength dependent SHG of flat and strain induced (-1.72%, -1.95% and -2.38%) wrinkled regions of 5L $WS_2$. SHG peak blue shifts with the compressive strain. **(d)** The band structure of 5L WS2 with the label of C calculated by the DFT. The arrows indicate the transition in A, B and the band nesting (C) **(e)** The wavelength dependent second order non-linear susceptibility ($I\chi$) spectra of 5L WS2 for three Strain (-2%, 0%, 2%) levels. Spectra are vertically shifted by 500nm and 1000nm for improved visibility. The C-exciton resonance enhanced SHG peak position is indicated by C. **(f)** The strain dependent SHG peak wavelength (in resonance with C exciton).

Black colour and red colour spheres represent calculated and experiments values respectively. Black dashed line is the linear fit of the calculated values.

In order to further explain the origin of this shift, we employed first-principles density functional theory (DFT) using simulation code Abinit to calculate the strain dependent wavelength dependent second order non-linear susceptibility $I\chi$ (-2ω, ω, ω) of 5L WS$_2$ (See Methods section for more details). Tensile strain causes red shifting of SHG peak whereas compressive strain results in blue shifting of SHG peak position. The simulation results show a remarkable shift in SHG peak position (in resonance with C-exciton) upon strain (%) as shown in **Figure 4e-4f** which is ~ 12.5 nm / Strain (%). The experimental measurements show a shift ~ 15 nm / Strain (%) which agrees well with the simulation results. Hence, wavelength dependent SHG clarifies the origin of resonance enhanced SHG peak shift in strain induced wrinkles thus provides another way to characterize fold and wrinkle nanostructures.

In conclusion, we have shown SHG as a sensitive and powerful tool to investigate the folding angle and strain angle accurately in 2D WS$_2$. Here, for the first time, we use polarization-dependent SHG technique to measure folding angle and strain vector in atomically thin tungsten disulfide (WS$_2$). Trilayer folds are found to show 9 times SHG enhancement due to the vector superposition of SH wave vectors coming from the individual layers of the fold with 60° folding angle. We find strain dependent SHG quenching and enhancement in the direction parallel and perpendicular respectively to the direction of the compressive strain vector. However, despite a variation in strain angle, total SHG remains constant which allow us to find the local strain vector using photoelastic approach. We also demonstrate that band-nesting induced transition (C peak) can highly enhance SHG, which can be significantly modulated by strain. Our results present an important advance, with applications in nonlinear optical devices.

**Methods**

**Buckled Sample Fabrication.** (i) WS$_2$ flakes are first exfoliated onto pre-buckled Gel-Film using scotch tape. (ii) Subsequently, the Gel film is released causing compressive forces on exfoliated WS$_2$ flakes generating well-aligned wrinkles perpendicular to the direction of the compressive forces; seem to cross the different layered samples. (iii-iv) The wrinkles fall down to form trilayer folds in 1-3L WS$_2$ whereas higher layered numbers (such as 5L) maintain their wrinkles' like curvature[50] (**Figure 1a**). Strained WS$_2$ samples are transferred on a Si/SiO$_{2/}$ substrate.[35]

**Experimental SHG setup**. We perform SHG measurements on Zeiss 780 Confocal Microscopy. The fundamental laser field is provided tunable pulse laser Ti:sapphire laser with a pulse width of 150 fs and a repetition rate of 80 MHz. A 50× confocal objective lens (NA = 0.85) is used to excite the sample. SHG measurements are taken on fundamental laser wavelength 900nm. The reflected SH signal is collected by the same objective, separated by a beam splitter and filtered by suitable optical filters to block the reflected fundamental radiation. The SH character of the detected radiation is verified by its wavelength and quadratic power dependence on the pump intensity. Laser with tunability range (800nm-1040nm) is used for wavelength dependent SHG. For polarization resolved SHG, an analyzer (polarizer) is used to select the polarization component of the SH radiation parallel to the polarization of the pump beam. The sample is rotated by a rotational stage to obtain the orientation dependence of the SH response.

**Simulations.** In the present work, the plane-wave method in the framework of DFT using Abinit code is employed. Local density approximation is used for the exchange-correlation effect, and the enery cutoff of 52 Ry is chosen for the electron wave function expansion. A vacuum layer thicker of more than 10 Å is included to avoid interaction between periodic layers. The *k*-point sampling is 24×24×1 for the prime cell of 5L WS$_2$.

## Author Contributions

Y. L. conceived and supervised the project; A. K., W. M prepared strained samples; A. K., B. L., A. S. carried out all the SHG measurements; A. K., Y. L. analyzed the data; A. K. carried out model calculations, L. Z., B. L. took the AFM imaging; A.K, W. M., Y. Z. helped with the schematic preparation; A. K. and Y. L. drafted the manuscript and all authors contributed to the manuscript.

## Acknowledgments

We would like to acknowledge the financial support from Australian National University. We would also like to acknowledge the support from Center of Advanced Microscopy (CAM), Australian National University.

## Conflict of Interest

The authors declare no competing financial interest.

## Supplementary Information

All additional data and supporting information are presented in the supplementary information file.

# Supplementary Information

# Direct Measurement of Folding Angle and Strain Vector in Atomically thin $WS_2$ using Second Harmonic Generation

Ahmed Raza Khan[1], Boqing Liu[1], Wendi Ma[1], Linglong Zhang[1], Ankur Sharma[1], Yi Zhu[1], Tieyu Lü[2] and Yuerui Lu[1*]

[1]Research School of Engineering, College of Engineering and Computer Science, Australian National University, Canberra ACT, 2601, Australia

[2]Department of Physics, and Institute of Theoretical Physics and Astrophysics, Xiamen University, Xiamen, 361005, China

\* To whom correspondence should be addressed: Yuerui Lu (yuerui.lu@anu.edu.au)

## S1. Strained samples fabrication

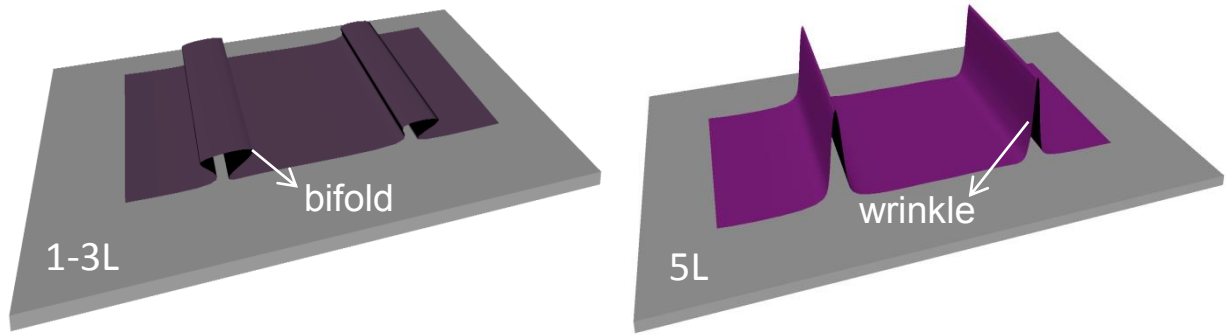

**Figure S1 | Description of strained samples fabrication.** (a) Bifold formation in 1-3L. (b) Wrinkle formation in >3L respectively.

$WS_2$ flakes are first exfoliated onto pre-buckled Gel-Film using scotch tape. Subsequently, the Gel film is released causing compressive forces on exfoliated $WS_2$ flakes generating well-aligned wrinkles perpendicular to the direction of the compressive forces; seem to cross the different layered samples. The wrinkles fall down to form bifolds in 1-3L $WS_2$ whereas higher layered numbers (such as 5L) maintain their wrinkles' like curvature.[1] (**Figure 1a**) An isometric view of folds (1-3L) and wrinkles (>3L) is shown in **Figure S1**. Strained $WS_2$ samples are transferred on a $Si/SiO_{2/}$ substrate.[2]

## S2. Power dependence on SHG intensity

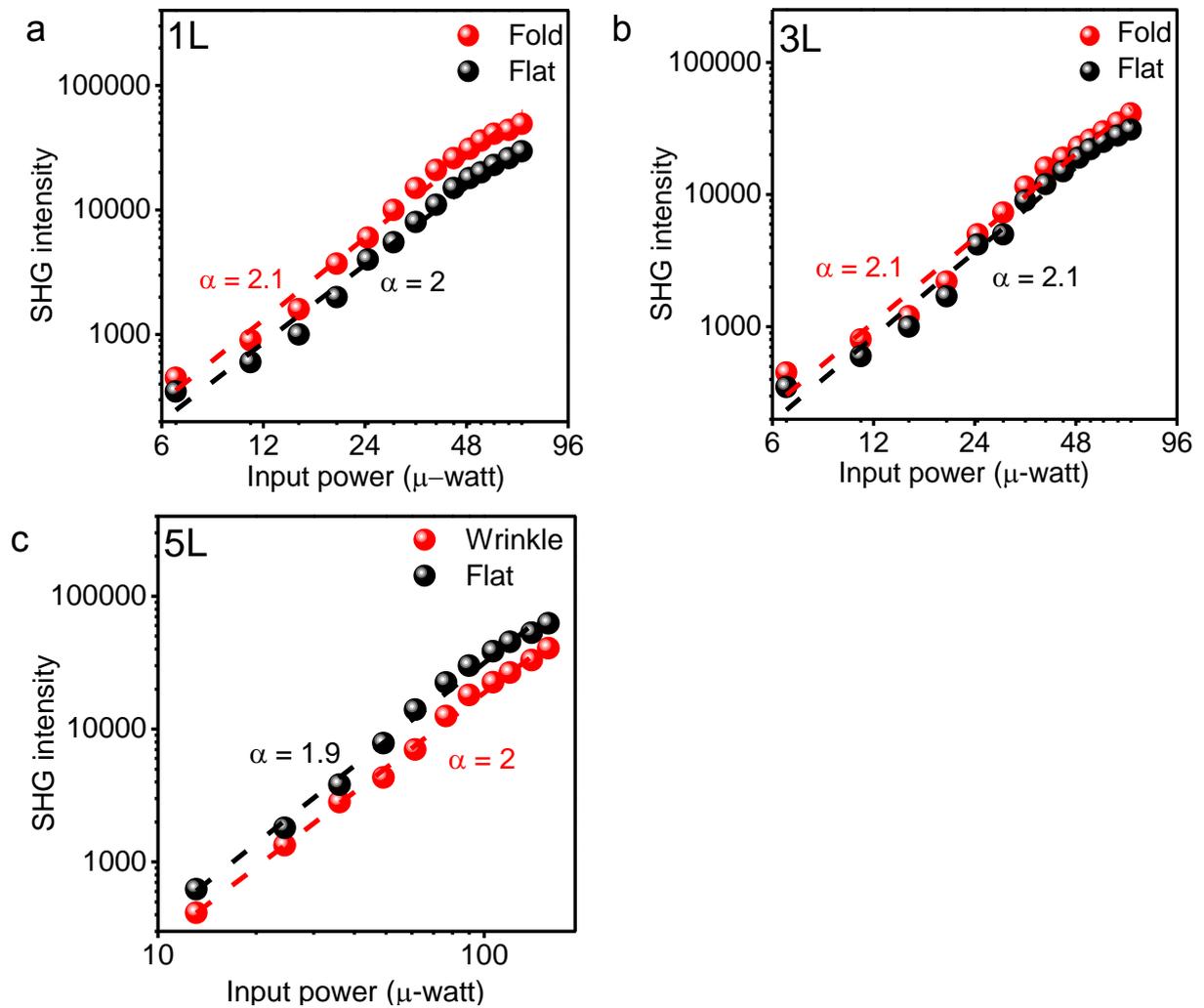

**Figure S2 | Power dependence on SHG intensity (a)** Power dependent SHG intensity at 900nm pump laser in a logarithmic scale of folded/wrinkled and flat regions of 1L. **(b)** 3L. **(c)** 5L. The measured SHG intensity is shown as points and the dashed lines are the linear fits of the data.

## S3. Layer dependent height of Folds/Wrinkles

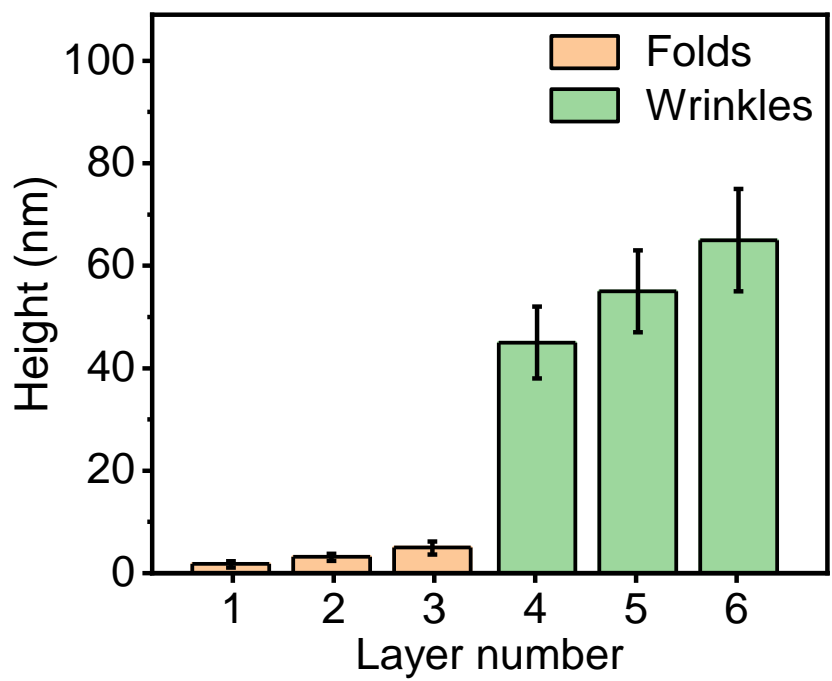

**Figure S3 | Layer dependent height of Folds (1-3L) and Wrinkles (5-6L)**

The height differences measured on wrinkle like protuberances in 1-3L buckled $WS_2$ are found to be 1.4 ± 0.5, 2.8 ± 0.5 and 4.2 ± 1 nm, respectively using Atomic Force Microscopy (AFM) measurements (**Figure S2**). These values match the height of 2L, 4L and 6L samples very well as the thickness of single layer is evaluated around 0.7 nm[3], which agrees well with the reported value. Therefore, wrinkles like protuberances in 1-3L can be regarded as the trilayer folds. In a previous study on strain engineering, the formation of folds is also observed in ultrathin $MoS_2$ due to collapse of wrinkles which are attributed to reduced elastic modulus of ultrathin layers.[1,4] AFM investigations of wrinkles in 4-6L $WS_2$ show a rapid increase in the height (~40-75nm) which shows that >3L wrinkles in $WS_2$ maintain their curvature.

## S4. Folding angle determination

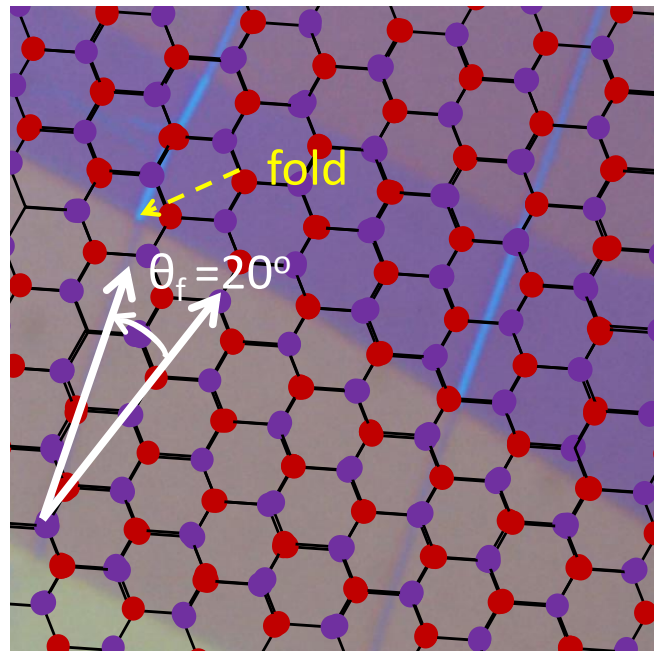

**Figure S4 | Folding angle determination.**

Folding angle ($\theta_f$) is the angle between AC (arm chair) direction and fold as shown in **Figure S4**. During parallel polarization SHG, maximum SHG signal is along AC direction.[5,6]

## S5. Strain estimation on wrinkle nanostructures

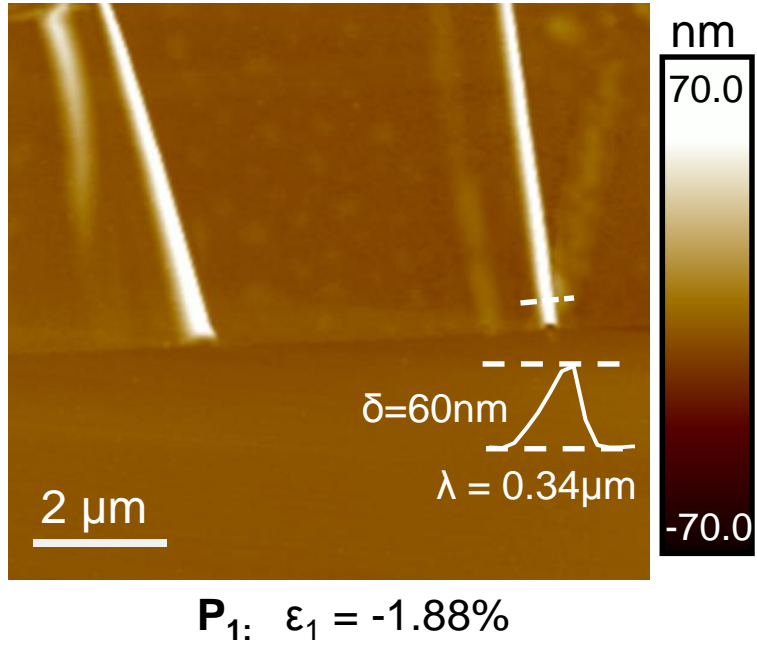

$P_1$:  $\varepsilon_1$ = -1.88%

**Figure S5 | Strain estimation on wrinkle.**

Maximum strain on the wrinkle is produced at the top of the wrinkles and it can be estimated as under;[1]

$$\varepsilon \sim \pi^2 h\delta/(1-v^2)\lambda^2 \qquad (2)$$

where $v$ is the Poisson's ratio, h is the thickness of the flake, and $\delta$ and $\lambda$ are the height and width of the wrinkle which were measured using atomic force microscopy (AFM). For $P_1$ point of the wrinkle $\delta = 60$nm, $\lambda = 340$nm, h = 3.5nm for 5L, $v = 0.22$ for $WS_2$[7], strain is calculated ~ -1.88%. Negative sign is used due to compressive strain.

## S6. Strain angle determination

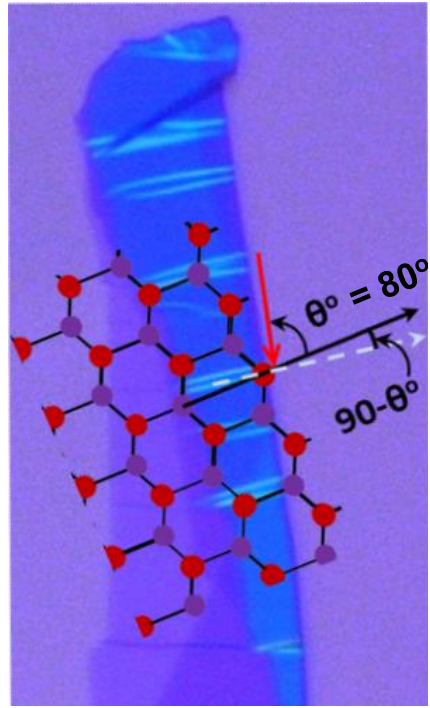

**Figure S6 | Strain angle determination.**

Strain angle ($\theta^o$) is the angle between strain direction (red arrow) and AC (arm chair) direction. Parallel SHG polar plot is used to find the AC direction (black arrow). The compressive strain direction (red arrow) is perpendicular to the wrinkles. Thus, strain angle ($\theta^o$) is calculated ~ 80°.

## S8. Wavelength dependent SHG of 3L fold

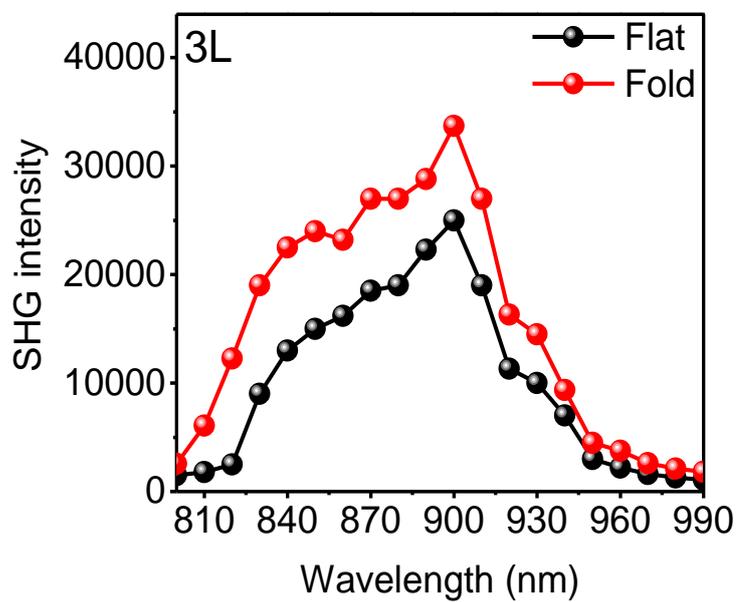

**Figure S8 | Wavelength dependent SHG of 3L fold**